%% file: sample-sigconf.tex
\documentclass[sigconf]{acmart}
\AtBeginDocument{%
  }

\copyrightyear{2026}
\acmYear{2026}
\setcopyright{cc}
\setcctype{by}
\acmConference[CIKM '26]{Proceedings of the 35th ACM International Conference on Information and Knowledge Management}{November 07--11, 2026}{Rome, Italy}
\acmBooktitle{Proceedings of the 35th ACM International Conference on Information and Knowledge Management (CIKM '26), November 07--11, 2026, Rome, Italy}
\acmDOI{10.1145/3799682.3840679}
\acmISBN{979-8-4007-2539-5/2026/11}
\usepackage{amsmath}
\usepackage{amsfonts}
\usepackage{multirow}
\usepackage{subfig}

\usepackage{times}
\usepackage{helvet}
\usepackage{courier}
\usepackage[dvipsnames]{xcolor}
\usepackage{tcolorbox}

\begin{document}

\title{On Mitigating Data Sparsity in Conversational Recommender Systems}

\author{Sixiao Zhang}
\affiliation{%
  \institution{Nanyang Technological University}
  \country{Singapore}
}
\email{sixiao001@e.ntu.edu.sg}

\author{Mingrui Liu}
\affiliation{%
  \institution{Nanyang Technological University}
  \country{Singapore}}
\email{mingrui001@e.ntu.edu.sg}

\author{Wei Yuan}
\affiliation{%
  \institution{The University of Queensland}
  \city{Brisbane}
   \state{Queensland}
  \country{Australia}
}
\email{w.yuan@uq.edu.au}

\author{Hongzhi Yin}
\authornote{Co-corresponding authors.}
\affiliation{%
 \institution{The University of Queensland}
 \city{Brisbane}
 \state{Queensland}
 \country{Australia}}
\email{h.yin1@uq.edu.au}

\author{Cheng Long}
\authornotemark[1]
\affiliation{%
  \institution{Nanyang Technological University}
  \country{Singapore}}
\email{c.long@ntu.edu.sg}

\renewcommand{\shortauthors}{Zhang et al.}

\begin{abstract}
    Conversational recommender systems (CRSs) infer user preferences from dialogue contexts, but they suffer from severe data sparsity in both dialogue and entity spaces. Dialogue data are linguistically diverse and open-ended, making it difficult to generalize across varied expressions. Meanwhile, existing CRS models often rely on large knowledge graphs, where only a small fraction of entities receive effective supervision during training, leaving the majority under-trained or entirely unseen at inference time. To address these challenges, we propose DACRS, a novel CRS framework consisting of three modules: Dialogue Augmentation, Knowledge-Guided Entity Modeling, and Dialogue–Entity Matching. The Dialogue Augmentation module adopts a two-stage augmentation pipeline to enrich dialogue contexts and improve robustness to linguistic variation. The Knowledge-Guided Entity Modeling module leverages knowledge graphs through entity substitution and an entity similarity constraint to enhance representation learning for sparsely supervised and unseen entities. Finally, the Dialogue–Entity Matching module integrates dialogue representations with mentioned entity embeddings via dialogue-guided attention aggregation, yielding user representations that capture both explicit and implicit preferences. Extensive experiments on two public benchmark datasets demonstrate that DACRS consistently outperforms state-of-the-art conversational recommender systems.
\end{abstract}

\ccsdesc[500]{Information systems~Information retrieval}
\ccsdesc[500]{Information systems~Recommender systems}
\ccsdesc[300]{Information systems~Retrieval models and ranking}
\ccsdesc[300]{Computing methodologies~Neural networks}
\keywords{Conversational recommender systems; data sparsity; data augmentation}


\maketitle

\input{introduction}
\input{related_works}
\input{methodology}
\input{experiments}

\section{Conclusion}

In this work, we investigate the data sparsity problem in conversational recommender systems, which includes both dialogue sparsity caused by the flexibility and redundancy of natural language, and item sparsity caused by long-tail distributions and sparse knowledge graph supervision. To address these challenges, we propose DACRS, a dual-path augmented conversational recommender framework consisting of three components: Dialogue Augmentation, Knowledge-Guided Entity Modeling, and Dialogue-Entity Matching. The Dialogue Augmentation module improves robustness to diverse dialogue expressions, the Knowledge-Guided Entity Modeling module enhances representation learning for sparsely supervised entities, and the Dialogue-Entity Matching module effectively integrates explicit semantic preferences with implicit collaborative signals for recommendation. Extensive experiments on two benchmark datasets demonstrate that DACRS consistently outperforms state-of-the-art conversational
recommender systems across multiple recommendation metrics. Further analyses, including ablation studies, hyperparameter studies, and embedding visualization, validate the effectiveness of each component and show that DACRS significantly improves entity
representation learning. Overall, our findings suggest that explicitly addressing both dialogue sparsity and item sparsity is crucial for building more robust and generalizable
conversational recommender systems.

\begin{acks}
This research is supported by the Ministry of Education, Singapore, under its Academic Research Fund (Tier 2 Award MOE-T2EP20224-0011 and Tier 1 Award (RG20/24)). Any opinions, findings and conclusions or recommendations expressed in this material are those of the author(s) and do not reflect the views of the Ministry of Education, Singapore. The Australian Research Council partially supports this work under the streams of Future Fellowship (Grant No. FT210100624), and the Linkage Project (Grant No. LP230200892, LP240200546 and LP250200778).
\end{acks}

\bibliographystyle{ACM-Reference-Format}
\balance
\bibliography{sample-base}










\end{document}

%% file: introduction.tex
\section{Introduction}

Conversational recommender systems (CRSs) \cite{sun2018conversational,ren2020crsal,ren2021learning,guo2023towards,zhang2026generative} provide an effective and real-time mechanism for capturing user preferences and delivering personalized recommendations. Unlike traditional recommender systems, CRSs allow users to interact through free-form natural language conversations, enabling them to express their needs more flexibly and intuitively. This interaction style offers a rich source of semantic information that can significantly enhance the accuracy and relevance of recommendations. Consequently, a core challenge in CRS development lies in how to best leverage the information embedded within these textual dialogues.

Current research on CRSs broadly follows two main directions. The first is the \textbf{LM-as-a-recommender} approach, where language models (LMs) process user dialogues directly to generate recommendations, either through reranking \cite{gao2023chat, huang2023recommender, li2024incorporating} or next-token generation \cite{fang2024multi, xi2024memocrs, he2024reindex}. However, LMs often lack domain-specific knowledge, making it difficult for them to accurately model user preferences and retrieve relevant items from a large candidate pool. As a result, these systems frequently underperform compared to \textbf{expert CRS models} \cite{wang2023recmind}, which is exactly the second direction. Expert CRS models represent users and items as embeddings and make recommendations based on similarity computations \cite{zhou2020improving, wang2022towards, yang2021improving}. These models consistently achieve state-of-the-art recommendation performance. The recommendations they generate can then be used to guide LLMs in producing coherent and contextually appropriate dialogue responses, making expert CRS models both more practical and effective in real-world applications.

\begin{figure}[t]
\centering
    \includegraphics[width=0.48\textwidth]{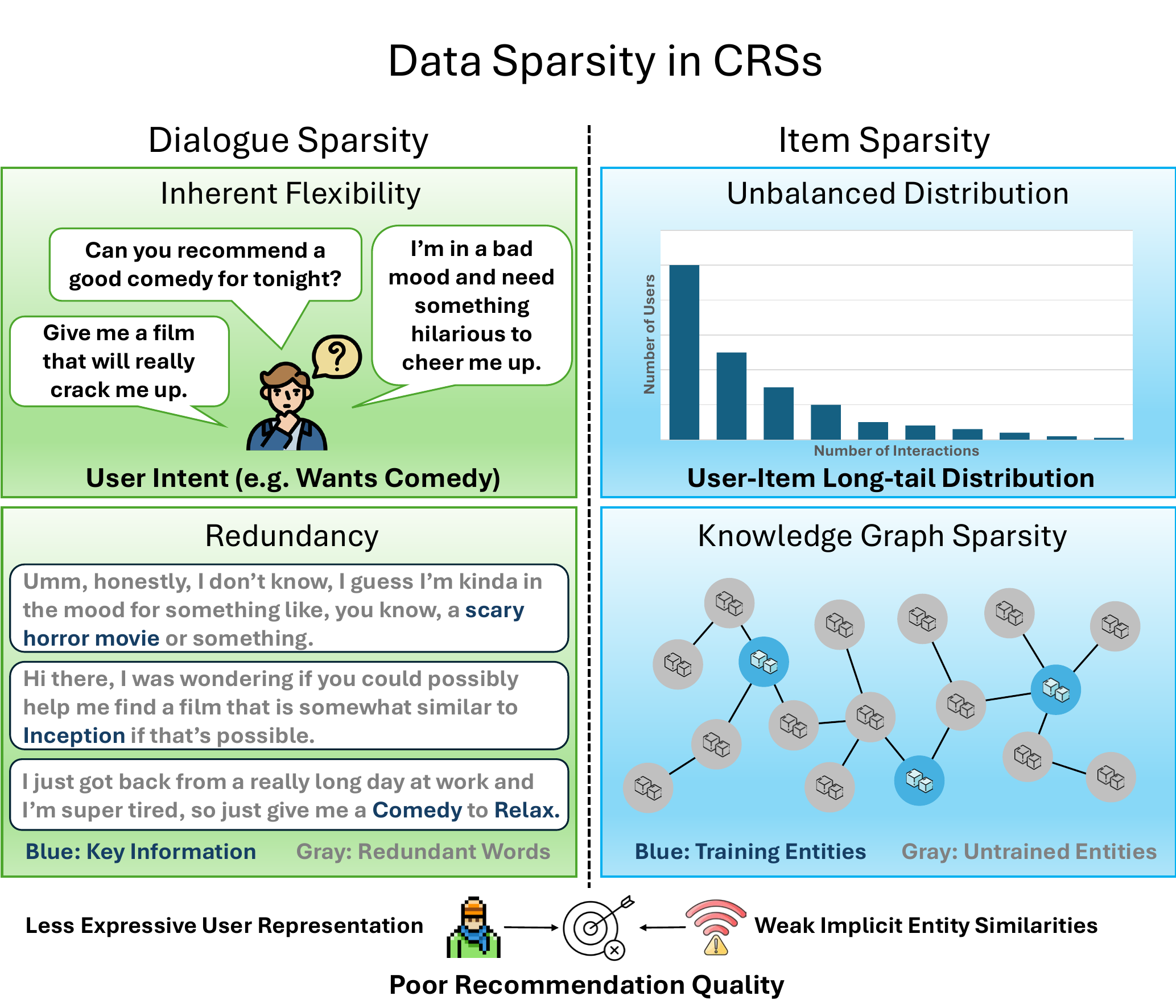}
\caption{Illustration of data sparsity in CRSs. The left panel depicts dialogue sparsity, arising from the high flexibility and redundancy of natural language expressions. The right panel illustrates item sparsity, characterized by highly unbalanced user–item interactions and sparse knowledge graphs containing a large number of entities that remain untrained.}
\label{fig:intro}
\end{figure}

Despite their advantages, existing approaches to CRSs often overlook the critical issue of data sparsity, which manifests in two forms: dialogue sparsity and item sparsity. An illustration of the data sparsity issue is shown in Figure~\ref{fig:intro}. On the one hand, dialogue modeling in CRSs faces two key challenges due to dialogue sparsity. First, language is inherently flexible, since identical preferences can be expressed in countless ways through varying phrasing, vocabulary, and structure. With limited training data, models often struggle to generalize to previously unseen dialogue variations \cite{belinkov2019analysis, li2020sentence}. Second, natural language is largely redundant \cite{bian2021attention}, and many tokens contribute noise rather than useful signal, introducing bias into the learned user embeddings. Most works underutilize the textual richness of user dialogues. They either extract only entities from conversations and discard the remaining text, or compress the entire dialogue into a single user embedding for item retrieval. Such strategies fail to capture the flexibility and nuance of natural language, leading to less expressive user representations.

On the other hand, item sparsity, which includes the long-tail distribution and the low overlap between entities in the knowledge graph (KG) and the dialogue corpus \cite{park2008long,abdollahpouri2017controlling}, poses a major obstacle for effective recommendation. Most items receive very few user interactions, which can lead to popularity bias, where models disproportionately favor popular items in their recommendations. For example, the user-item interaction density is just 0.0956\% in the ReDial dataset and 0.2637\% in the Inspired dataset. To mitigate this, many systems \cite{chen2019towards,zhou2020improving,wang2022towards,li2024incorporating} incorporate a KG composed of items and related entities to establish stronger correlations among items. However, these graphs are often underutilized due to a limited overlap with the training corpus. For example, ReDial’s KG includes 31,162 entities, yet only 6,675 of them appear in the training corpus. Meanwhile, there are 2,129 entities in the test corpus, but 415 of them never appear in the training corpus. This disconnect exacerbates the data sparsity issue for items, as most KG entities are rarely or never updated during training. Consequently, neighborhood aggregation fails to effectively capture implicit entity similarities, limiting the recommender system’s generalization ability.

To this end, we aim to address these problems in this paper. We propose a model named \textbf{D}ual-path \textbf{A}ugmented \textbf{C}onversational \textbf{R}ecommender \textbf{S}ystem (DACRS). It consists of three components: \textbf{Dialogue Augmentation}, \textbf{Knowledge-Guided Entity Modeling}, and \textbf{Dialogue-Entity Matching}. Specifically, \textbf{Dialog Augmentation} aims to capture the explicit user semantic preference (e.g., funny, supernatural) while improving the model's generalization ability to varying texts. We adopt a two-stage augmentation pipeline to generate augmented texts by harnessing the superior language understanding and generation ability of LLMs. In the first stage, we rewrite the dialogue through LLM rephrasing and LLM summarization to increase the diversity of the texts and increase the generalizability of the model to unseen dialogues. In the second stage, we apply word-level and utterance-level augmentations to mitigate the noise of redundant language. \textbf{Knowledge-Guided Entity Modeling} aims to capture the implicit user preference through mentioned entities while addressing the sparsity problem to learn stronger item representations. We alter the mentioned entities with their 1-hop neighbors from the KG to enhance local generalization. A novel entity similarity constraint is proposed to encourage high similarity between connected entities. \textbf{Dialogue-Entity Matching} encodes entities and items through RGCN \cite{schlichtkrull2018modeling}, and encodes dialogues through LLMs. The dialogue embedding is fused with the embeddings of the mentioned entities through a dialogue-guided attention aggregation module to capture both user explicit and implicit preferences. We compute similarity scores using user and item embeddings and make recommendations with top-ranked items. 
We release our code for reproduction\footnote{https://github.com/RinneSz/DACRS-Dual-path-Augmentation-for-Conversational-Recommender-Systems}.

We summarize our contributions as follows:
\begin{itemize}
    \item We address the data sparsity problem in CRSs, which includes dialogue sparsity (inherent flexibility and redundancy) and item sparsity (unbalanced distribution and knowledge graph sparsity).
    \item We propose DACRS that consists of three modules: Dialogue Augmentation, Knowledge-Guided Entity Modeling, and Dialogue-Entity Matching to address the sparsity issues.
    \item We show the state-of-the-art performance of DACRS on two popular datasets. Ablation study, hyperparameter study, and visualization are provided to validate the effectiveness of each component.
\end{itemize}

%% file: related_works.tex
\section{Related Works}
Existing works on CRSs can be broadly categorized into LM-as-a-recommender models and expert CRS models. In this section, we review these two lines of research.

\subsection{LM-as-a-recommender}
LM-as-a-recommender approaches directly leverage language models to perform recommendation, typically via reranking or next-token generation. \cite{li2018towards} employs a GRU-based text generator where items are embedded as text tokens, but it can only recommend items explicitly mentioned in the dialogue. KBRD~\cite{chen2019towards} derives user embeddings by mean-pooling mentioned entities from the knowledge graph, which are then transformed into a bias vector added to the attention layer of a dialogue transformer; both responses and recommendations are generated in an end-to-end manner. \cite{zhou2020towards} predicts the next conversation topic and prompts a GPT model with both the dialogue history and the predicted topic to generate responses and recommendations. Chat-Rec~\cite{gao2023chat} constructs prompts from dialogue content and user profiles to query ChatGPT; if recommendation is required, an external recommender produces top-$k$ candidates, which are subsequently reranked by the LLM. InteRecAgent~\cite{huang2023recommender} instructs LLMs to generate tool-calling plans based on the dialogue and executes these plans to retrieve and rerank candidate items. LLMCRS~\cite{feng2023large} designs a multi-stage system comprising sub-task detection, model matching, sub-task execution, and response generation, all realized through prompting LLMs. ChatCRS~\cite{li2024incorporating} prompts an LLM to predict conversation goals while retrieving KG neighbors of mentioned entities, and then generates responses conditioned on both sources of information. CRAG~\cite{zhu2025collaborative} first retrieves candidate items using traditional collaborative filtering methods and subsequently relies on LLMs for reranking. \cite{surana2025reviews} synthesizes user dialogues to mitigate low-resource settings; however, the use of fixed templates and in-corpus items leaves sparsity issues largely unresolved. These methods effectively integrate LLMs into the recommendation pipeline and generate fluent, human-like responses. Nevertheless, most LM-as-a-recommender approaches struggle to outperform expert CRS models, as LLMs often fail to capture task-specific implicit user preferences~\cite{wang2023recmind}.

\subsection{Expert CRS Models}
Expert CRS models explicitly learn latent representations of users and items and perform recommendation by measuring similarity in the embedding space. KGSF \cite{zhou2020improving} learns word embeddings from a word graph and entity embeddings from an entity KG, and fuses the two spaces via mutual information maximization to compute recommendation scores. MESE \cite{yang2021improving} encodes item metadata directly as item embeddings and fine-tunes GPT with special tokens to model conversational patterns; recommendations are generated by matching item embeddings with the REC token embedding. UniCRS \cite{wang2022towards} computes dialogue word embeddings using an LLM and entity embeddings using RGCN, aligns the two spaces through a bilinear transformation, and queries an LLM with fused embeddings and task-specific latent prompts. DCRS \cite{dao2024broadening} retrieves semantically similar dialogues as in-context demonstrations to enhance conversational recommendation through demonstration-augmented prompt learning. STEP \cite{yang2025step} adopts a curriculum learning strategy to progressively align dialogue semantics with knowledge graph entities for improved recommendation. MSCRS \cite{wei2025mscrs} incorporates multimodal information to construct modality-specific graphs and integrates them with prompt learning for conversational recommendation. ReFICR \cite{yang2024unleashing} fine-tunes an LLM to perform CRS without an external recommendation module by encoding item metadata and dialogue texts into embeddings, retrieving candidate items via cosine similarity, and prompting the LLM for reranking. COLA \cite{lin2023cola} augments the entity KG with a user-item interaction graph to obtain interaction-aware embeddings and performs recommendation based on user-item similarities. CFCRS \cite{wang2023improving} enlarges the training data by augmenting dialogues with entities from other conversations. These expert CRS models achieve state-of-the-art recommendation performance by more effectively capturing implicit user preferences. However, they often overlook language diversity and item sparsity, which limits their generalization ability under severe sparsity.

%% file: methodology.tex
\begin{figure*}[t]
\centering
    \includegraphics[width=1.0\textwidth]{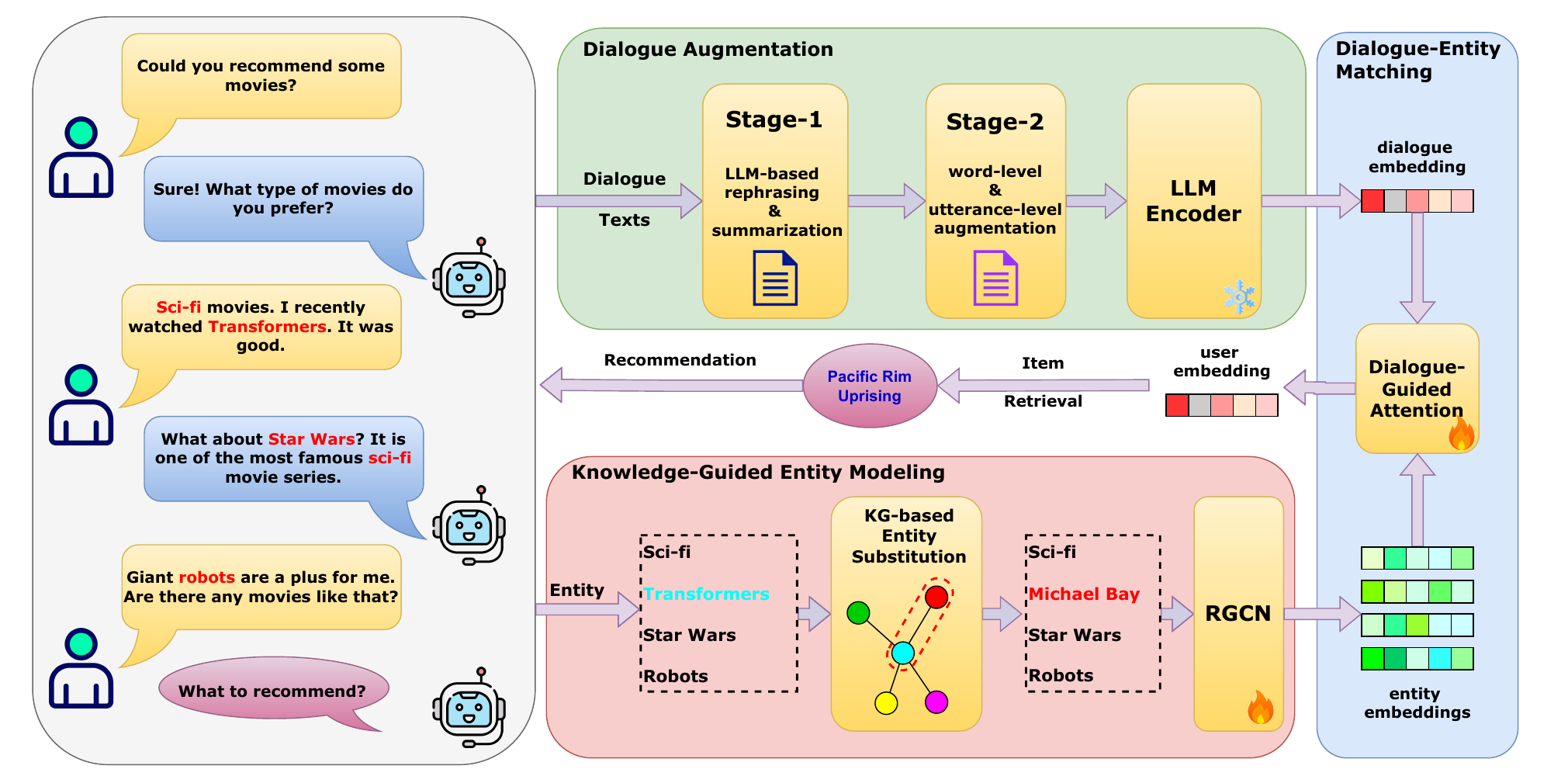}
\vspace{-2em}
\caption{An illustration of DACRS. It consists of three modules. (1) \textbf{Dialogue Augmentation}. The dialogue goes through a two-stage augmentation pipeline and obtain the dialogue embedding through an LLM encoder. (2) \textbf{Knowledge-Guided Entity Modeling}. The mentioned entities are replaced with their 1-hop neighbor in the knowledge graph encoded through an RGCN. (3) \textbf{Dialogue-Entity Matching}. The dialogue embedding and the entity embeddings are fused through a dialogue-guided attention network to obtain the user embedding and retrieve recommendations.}
\label{fig:method}
\vspace{-1em}
\end{figure*}

\section{Dual-path Augmentation for Conversational Recommender Systems}
We introduce our \textbf{D}ual-path \textbf{A}ugmentation for \textbf{C}onversational \textbf{R}-ecommender \textbf{S}ystems (DACRS). It consists of three modules, namely \textbf{Dialogue Augmentation}, \textbf{Knowledge-Guided Entity Modeling}, and \textbf{Dialogue-Entity Matching}. An illustration of DACRS is shown in Figure \ref{fig:method}. The \textbf{Dialogue Augmentation} module aims to capture user's explicit semantic preference from the dialogue (e.g., robots, funny, supernatural). We address the challenges of flexible and nuance language by introducing a two-stage dialogue augmentation pipeline. The \textbf{Knowledge-Guided Entity Modeling} module captures the implicit user preference through the mentioned entities. We alleviate the sparsity problem of entities and the knowledge graph by KG-based entity substitution and an entity similarity constraint. The \textbf{Dialogue-Entity Matching} module encodes the entities and dialogues into latent embeddings, and obtain the user embedding by fusing both embeddings through a dialogue-guided attention network.

\subsection{Problem Definition}
\label{sec:problem definition}

Each dialogue is conducted by a unique user interacting with the recommender. Let the set of all dialogues be denoted as $\mathcal{U} = \{\mathcal{U}_1, \mathcal{U}_2, \ldots, \mathcal{U}_N\}$, where $N$ is the number of users. A dialogue $\mathcal{U}_n$ consists of a sequence of utterances $\mathcal{U}_n = \{u_n^1, u_n^2, \ldots, u_n^{|\mathcal{U}_n|}\}$, where each utterance is produced either by the user or the recommender. Given a set of candidate items $\mathcal{I}=\{i_1,i_2,\ldots,i_I\}$, the goal of a CRS is to recommend items that align with the user’s preferences in the recommender’s next utterance $u_n^{|\mathcal{U}_n|+1}$.

Relying solely on dialogue context is insufficient for accurate recommendations, particularly without incorporating knowledge of candidate items. To address this, we leverage an item-entity knowledge graph to capture item correlations. For each dialogue $\mathcal{U}_n$, we extract the set of mentioned entities $\mathcal{E}_n = \{e_n^1, e_n^2, \ldots, e_n^{|\mathcal{E}_n|}\}$, following prior works \cite{zhou2020improving, wang2022towards}. The KG, denoted as $\mathcal{G}$, is constructed using DBpedia \cite{bizer2009dbpedia}, where each node represents an entity (including items), and edges represent typed relations between them. For instance, the triplet \{$\textit{Chris Evans} - \textit{starring} - \textit{Captain America}$\} indicates a relation in the KG, with nodes connected by the edge \textit{starring}. With this setup, the CRS task is formalized as: 
\begin{equation} 
\mathcal{I}_n = f(\mathcal{U}_n, \mathcal{E}_n, \mathcal{G})
\end{equation} 
where $f$ denotes the CRS model, and $\mathcal{I}_n = \{i_n^1, i_n^2, \ldots, i_n^k\}$ represents the top-k recommended items for dialogue $\mathcal{U}_n$.

\subsection{Dialogue Augmentation Module}
This module aims to capture the explicit semantic preference of the user from the dialogue. We propose a two-stage dialogue augmentation pipeline designed to generate semantically consistent variations of dialogues, thereby enriching the training data and enhancing model generalizability. In Stage 1, we obtain an altered text $\mathcal{U}_n'$ by employing a large language model (LLM) to rephrase or summarize the dialogue, generating alternate expressions that preserve user intent. In Stage 2, we apply a set of text augmentation techniques on $\mathcal{U}_n'$ to further introduce diversity and reduce the impact of redundant tokens to obtain $\mathcal{U}_n''$. Each raw dialogue is sequentially processed through both stages during training, with one augmentation method applied per stage. Note that we also set a probability for each stage to output the original text without augmentation. 

\subsubsection{Stage 1: LLM-Based Rephrasing \& Summarization}
In the first stage, we apply LLM-based rephrasing and summarization to enhance dialogue diversity.
\paragraph{Rephrasing.}
Given a dialogue, we prompt an LLM (e.g., Llama, GPT) to rephrase the conversation. The prompt is:

\begin{tcolorbox}[title=Rephrasing Prompt]
    You are given a dialogue between a user and a recommender system. Here is the dialogue: \{dialogue\}. Rephrase the dialogue using as different words and styles as possible. Output ONLY the rephrased content.
\end{tcolorbox}

The rephrased dialogue differs from the original in vocabulary and expression style, while preserving the user’s underlying preferences and maintaining the interactive format of a conversation. This augmentation leverages the strong generative capabilities of LLMs to create diverse linguistic variations of the same semantic content. As a result, the model is exposed to varied surface forms expressing the same user intent, improving its generalization to unseen dialogue expressions.

\paragraph{Summarization.}
Alternatively, we prompt the LLM to summarize the user’s preferences expressed in the dialogue. This produces a concise representation that distills the user’s intent into an informative summary. The prompt is:

\begin{tcolorbox}[title=Summarization Prompt]
You are given a dialogue between a user and a recommender system. Here is the dialogue: \{dialogue\}. Summarize the user's preference for movie recommendations in a compact and informative manner.
\end{tcolorbox}

Unlike the rephrased dialogue, the summary does not retain the conversational format. Instead, it generates a direct and structured statement of the user’s preferences using explicit concepts. For instance, the summary “the user prefers sci-fi movies with giant robots” could represent the preference expressed in the dialogue shown in Figure \ref{fig:method}. This variant provides the model with clear and focused signals, helping it attend to core preference information while ignoring irrelevant or noisy dialogue content.

\subsubsection{Stage 2: Text Augmentation}
In the second stage, we apply both word-level and utterance-level augmentation techniques to the outputs from Stage 1. These augmentations mitigate redundancy in natural language and expose the model to a wider distribution of text patterns. The augmentation methods include:

\subsubsection*{Word-Level Augmentations}
\begin{itemize}
    \item \textbf{Deletion}: Randomly remove a proportion of words.
    \item \textbf{Swapping}: Swap adjacent words to simulate alternate phrasing.
    \item \textbf{Cropping}: Remove a contiguous segment of words.
\end{itemize}

\subsubsection*{Utterance-Level Augmentations}
\begin{itemize}
    \item \textbf{Deletion}: Remove one or more utterances.
    \item \textbf{Swapping}: Swap the positions of two utterances.
\end{itemize}
If summarization is used in Stage 1, only word-level augmentations are applied in Stage 2, as the dialogue no longer contains distinct utterances. These augmentations create diverse inputs with similar user semantic preferences, enabling the model to learn robust, semantically grounded dialogue representations and generalize better to real-world dialogue variability.

\subsection{Knowledge-Guided Entity Modeling Module}
This module aims to capture implicit user preferences from the mentioned entities. Similar to conventional recommendation tasks, conversational recommender systems also suffer from long-tail distributions and severe data sparsity. However, a key distinction between CRS datasets and traditional recommendation datasets lies in the presence of a large number of unseen items and entities in the knowledge graph. For instance, in the ReDial dataset, only 6,675 out of 31,162 entities appear in the training corpus, and 415 out of the 2,129 entities in the test set never occur during training. As a result, rarely mentioned items are often weakly connected to frequently mentioned ones in the KG, which limits the effectiveness of relational information for learning informative item representations. To mitigate this issue, we introduce two complementary components to enhance item representation learning: KG-based entity substitution and an entity similarity constraint.

\subsubsection{KG-Based Entity Substitution}

We observe that most neighboring entities in the KG are not mentioned in the dialogue. For example, in the ReDial dataset, the average node degree is 4.84, yet only 2.47 of these neighbors are mentioned on average. Similarly, in the Inspired dataset, the average node degree is 3.33, while only 1.32 neighbors are mentioned. To strengthen inter-entity connections, we introduce a substitution strategy that replaces mentioned entities with their 1-hop neighbors in the KG. Specifically, for each dialogue $\mathcal{U}_n$, we extract the set of mentioned entities $\mathcal{E}_n = {e_n^{1}, e_n^{2}, \ldots, e_n^{|\mathcal{E}_n|}}$ and randomly substitute a subset of them with their 1-hop neighbors. As a result, nearly twice as many entities are involved during training, and this number further increases as the graph convolutional network propagates training signals to neighboring nodes. This strategy introduces a richer diversity of entities into the training process and encourages semantically connected entities to share similar preference representations.

\subsubsection{Entity Similarity Constraint}

While KG-based substitution enriches training signals, it does not guarantee that all entities, especially those multiple hops away, are effectively trained. To further address this, we introduce an entity similarity constraint that encourages neighboring entities in the KG to have similar embeddings. We define it as follows: 
\begin{equation} 
\mathcal{L}_{\text{entity}, m} = -\sum_{j \in \mathcal{N}_m} \log \frac{e^{\mathbf{h}_m \cdot \mathbf{h}_j}}{\sum_{k=1}^{|\mathcal{E}|} e^{\mathbf{h}_m \cdot \mathbf{h}_k}} 
\end{equation} 

\begin{equation} 
\label{eq:kg loss} 
\mathcal{L}_{\text{entity}} = \sum_{m=1}^{|\mathcal{E}|} \mathcal{L}_{\text{entity}, m} 
\end{equation} 
where $\mathcal{N}_m$ denotes the set of neighbors of entity $m$ in the KG, $|\mathcal{E}|$ is the total number of entities, and $\mathbf{h}_m$ is the embedding of entity $m$. This constraint encourages connected entities to be closer in the embedding space than unconnected ones. As a result, even entities that do not appear in the training data can still learn meaningful embeddings through their structural connections in the KG.

\subsection{Dialogue-Entity Matching Module}
In this module, we compute personalized recommendations by matching dialogue and entity embeddings. Specifically, we first obtain entity embeddings using a Relational Graph Convolutional Network (RGCN) \cite{schlichtkrull2018modeling} and derive dialogue embeddings via a fixed LLM encoder. To represent user preferences, we perform a dialogue-guided attention aggregation on the embeddings of mentioned entities. Recommendations are then generated based on the similarity between user and item embeddings.

\subsubsection{Entity Encoding}
Following prior works \cite{zhou2020improving,wang2022towards}, we use RGCN \cite{schlichtkrull2018modeling} to encode entity and item embeddings from the knowledge graph. The RGCN layer is defined as:
\begin{equation}
    \mathbf{h}_m^{(l+1)} = \sigma \left( \sum_{r \in \mathcal{R}} \sum_{j \in \mathcal{N}_m^r} \frac{1}{c_{m,r}} \mathbf{W}_r^{(l)} \mathbf{h}_j^{(l)} + \mathbf{W}_0^{(l)} \mathbf{h}_m^{(l)} \right)
\end{equation}
where $\sigma$ is a non-linear activation function, $\mathcal{R}$ is the set of relations, $\mathcal{N}_m^r$ is the set of neighbors of entity $m$ under relation $r$, and $c_{m,r}$ is a degree-based normalization constant. $\mathbf{W}_r^{(l)}$ and $\mathbf{W}_0^{(l)}$ are learnable weight matrices, and $\mathbf{h}_m^{(l)}$ is the representation of entity $m$ at layer $l$. The initial embeddings $\mathbf{h}^{(0)}$ are randomly initialized and optimized during training. For brevity, we omit the layer index in subsequent discussion and denote the final embedding of entity $m$ as $\mathbf{h}_m$.

\subsubsection{User Encoding via Dialogue-Guided Attention Aggregation}
After augmenting a user dialogue $\mathcal{U}_n$ using the dialogue augmentation module, we obtain the modified input $\mathcal{U}_n''$. We then pass $\mathcal{U}_n''$ through a fixed LLM to produce a dialogue embedding:
\begin{equation}
    \mathbf{s}_n=\text{LLM}(\mathcal{U}_n'')
\end{equation}
The resulting embedding $\mathbf{s}_n$ captures the user’s explicit semantic preferences. However, general-purpose LLMs lack task-specific knowledge and are unaware of candidate items, limiting their ability to model implicit preferences. To mitigate this, we extract the set of mentioned entities $\mathcal{E}_n = \{e_n^1, e_n^2, \ldots, e_n^{|\mathcal{E}_n|}\}$ from the dialogue, and obtain their embeddings $\mathbf{H}_n \in \mathbb{R}^{|\mathcal{E}_n| \times d}$ from the RGCN. We then apply a dialogue-guided attention to aggregate the entity embeddings under the guidance of the dialogue embedding:

\begin{equation}
    Q_n = \textbf{s}_n W_Q;\ K_n=\textbf{H}_n W_K;\ V_n=\textbf{H}_n W_V
\end{equation}
  
\begin{equation}
    \mathbf{h}_n=\text{softmax}\left( \frac{Q_n K_n^\top}{\sqrt{d}} \right) V_n
\end{equation}
Here, $W_Q \in \mathbb{R}^{d \times d_{\text{llm}}}$ projects the LLM embedding into the same dimension as the entity embeddings. $W_K, W_V \in \mathbb{R}^{d \times d}$ are learnable key and value projection matrices. We omit the multi-head design for simplicity. The aggregated vector $\mathbf{h}_n$ captures the user’s implicit preferences by leveraging collaborative signals from semantically related entities. To form the final user embedding $\mathbf{u}_n$, we combine both explicit and implicit representations with a learnable weighted aggregation:
\begin{equation}
    \mathbf{u}_n=\lambda\mathbf{s}_n+(1-\lambda)\mathbf{h}_n
\end{equation}
\begin{equation}
   \lambda = \sigma(\mathbf{w}^\top\left[ \mathbf{s}_n \,\|\, \mathbf{h}_n \right]+b)
\end{equation}
where $\lambda$ is a scalar in (0,1), $\sigma(\cdot)$ is the sigmoid function, $[\cdot\|\cdot]$ is the concatenation function, $\mathbf{w}\in\mathbb{R}^{2d}$ and $b\in\mathbb{R}$ are the learnable weight and bias. This fusion allows $\mathbf{u}_n$ to encode both semantic information from dialogue context and collaborative signals from the mentioned entities, facilitating more informed recommendation decisions.

\subsection{Training Objective}
To train the model, we minimize the cross-entropy loss between users and their mentioned entities. For user $n$, the loss is defined as:
\begin{equation}
    \mathcal{L}_{n}=-\sum_{j\in\mathcal{E}_n}\text{log}\frac{e^{\mathbf{u}_n\cdot\mathbf{e}_j}}{\sum_{k=1}^{|\mathcal{E}|}e^{\mathbf{u}_n\cdot\mathbf{e}_k}}\\
\end{equation}
where $\mathcal{E}_n$ is the set of mentioned entities in the dialogue, $|\mathcal{E}|$ is the total number of entities, and $\mathbf{e}_j$ is the embedding of entity $j$. The overall recommendation loss is:
\begin{equation}
\label{eq:rec loss}
    \mathcal{L}_{\text{rec}}=\sum_{n=1}^{N}\mathcal{L}_n
\end{equation}
where $N$ is the total number of users. The final training objective combines the recommendation loss with the entity similarity constraint in Equation \ref{eq:kg loss} with a hyperparameter $\alpha$ adjusting the weight of the entity similarity constraint:
\begin{equation}
\label{eq:final loss}
    \mathcal{L}=\mathcal{L}_{\text{rec}}+\alpha\mathcal{L}_{\text{entity}}
\end{equation}

\subsection{Inference Procedure}

At inference time, given a user representation $\mathbf{u}_n$ and an item representation $\mathbf{h}_m$, we compute their relevance score using a dot-product similarity function:
\begin{equation}
    \text{score}_{n,m} = \mathbf{u}_n \cdot \mathbf{h}_m^\top.
\end{equation}
This formulation enables efficient similarity computation while remaining consistent with common practice in embedding-based recommendation systems.

For each user $n$, the model computes similarity scores against all candidate item entities in the item embedding space. It is worth emphasizing that only \emph{item entities} are involved in the final recommendation stage, while non-item entities are excluded from ranking. Based on the computed scores, all candidate items are sorted in descending order, and the top-$k$ items with the highest scores are returned as the final recommendation list.

\subsection{Efficiency and Complexity Analysis}

From an efficiency perspective, our method introduces no additional inference-time computational overhead compared to standard embedding-based recommendation models. The proposed Dialogue Augmentation module and Knowledge-guided Entity Modeling module are exclusively activated during the training phase to enhance representation learning. During inference, these modules are entirely disabled, and no synthetic dialogues or auxiliary knowledge signals are introduced.

Consequently, inference relies solely on the Dialogue-entity Matching module, whose inputs consist of (1) the embedding of the raw dialogue context and (2) the embeddings of the entities explicitly mentioned in the dialogue. This design ensures that inference remains lightweight and deterministic, making the proposed framework readily applicable to large-scale, real-world deployment scenarios where latency and computational efficiency are critical concerns.

%% file: experiments.tex
\begin{table}[ht]
\setlength{\tabcolsep}{10pt}
\caption{Dataset Statistics.}
\begin{center}
 \begin{tabular}{cccc} 
 \hline
  & ReDial & Inspired  \\
 \hline
 \#Dialogues & 11,348 & 999 \\
 \#Items & 6,281 & 1,472 \\
 \#Entities & 31,162 & 17,321 \\
 \#Interactions & 60,072 & 3,878 \\
 \hline
\end{tabular}
\label{table:datasets}
\end{center}
\end{table}

\section{Experiments}
We aim to answer these research questions:
\begin{itemize}
    \item \textbf{RQ1}: How does DACRS perform in terms of recommendation accuracy?
    \item \textbf{RQ2}: How does each module contribute to the performance of DACRS?
    \item \textbf{RQ3}: How do the hyperparameters influence the performance of DACRS?
    \item \textbf{RQ4}: Can DACRS effectively learn representations for entities that never appear in the training data?
\end{itemize}
We first introduce the datasets, evaluation metrics, baselines, and
experimental settings, followed by analyses corresponding to each
research question.

\subsection{Datasets}

We conduct experiments on two widely used English conversational
recommendation benchmarks, namely ReDial
\cite{li2018towards} and Inspired \cite{hayati2020inspired}.
Both datasets are collected through crowd-sourced conversations
between a seeker and a recommender. The statistics of the datasets
are summarized in \autoref{table:datasets}. Following previous
works \cite{chen2019towards,zhou2020improving,wang2022towards},
we construct the knowledge graph using DBpedia \cite{bizer2009dbpedia}, where items and their related entities are
treated as graph nodes. For ReDial, 10,006 dialogues are used for training and 1,342 for testing. For Inspired, 900 dialogues are used for training and 99 for testing.

\begin{table*}[t]
\begin{center}
 \caption{Recommendation performance on ReDial. Boldfaced numbers are the best, and underlined numbers are the second. $\ast$ denotes statistic significance (p-value < 0.05) against the best baseline.}
\begin{tabular}{|l|l|l|l|l|l|l|l|l|l|l|l|l|}
\hline
Method & R@5 & R@10 & R@20 & R@50 & N@5 & N@10 & N@20 & N@50 & M@5 & M@10 & M@20 & M@50 \\
\hline
KGSF & 0.119 & 0.179 & 0.262 & 0.364 & 0.076 & 0.096 & 0.117 & 0.137 & 0.062 & 0.071 & 0.076 & 0.079 \\
MESE & 0.135 & 0.210 & 0.294 & 0.413 & 0.088 & 0.112 & 0.134 & 0.157 & 0.073 & 0.083 & 0.089 & 0.092 \\
UniCRS & 0.149 & 0.215 & 0.296 & 0.418 & 0.098 & 0.119 & 0.140 & 0.164 & 0.082 & 0.090 & 0.096 & 0.100 \\
DCRS & 0.152 & 0.219 & 0.297 & 0.411 & 0.102 & 0.124 & 0.143 & 0.166 & 0.086 & 0.095 & 0.100 & 0.104 \\
STEP & 0.153 & 0.224 & \underline{0.309} & \underline{0.419} & 0.106 & 0.128 & \underline{0.150} & \underline{0.172} & 0.090 & 0.099 & 0.105 & 0.109 \\
MSCRS & \underline{0.160} & \underline{0.230} & 0.305 & 0.417 & \underline{0.108} & \underline{0.131} & \underline{0.150} & \underline{0.172} & \underline{0.091} & \underline{0.101} & \underline{0.106} & \underline{0.110} \\
DACRS & \textbf{0.164} & $\textbf{0.255}^{*}$ & $\textbf{0.335}^{*}$ & $\textbf{0.467}^{*}$ & \textbf{0.110} & $\textbf{0.138}^{*}$ & \textbf{0.154} & \textbf{0.175} & \textbf{0.095} & $\textbf{0.110}^{*}$ & \textbf{0.108} & \textbf{0.113} \\
\hline
Improv. & \textcolor{ForestGreen}{2.5\%} & \textcolor{ForestGreen}{10.87\%} & \textcolor{ForestGreen}{8.41\%} & \textcolor{ForestGreen}{11.46\%} & \textcolor{ForestGreen}{1.85\%} & \textcolor{ForestGreen}{5.34\%} & \textcolor{ForestGreen}{2.67\%} & \textcolor{ForestGreen}{1.74\%} & \textcolor{ForestGreen}{4.40\%} & \textcolor{ForestGreen}{8.91\%} & \textcolor{ForestGreen}{1.89\%} & \textcolor{ForestGreen}{2.73\%} \\
\hline
\end{tabular}
\label{table:redial rec}
\end{center}
\end{table*}

\begin{table*}[t]
\begin{center}
 \caption{Recommendation performance on Inspired. Boldfaced numbers are the best, and underlined numbers are the second. $\ast$ denotes statistic significance (p-value < 0.05) against the best baseline.}
\begin{tabular}{|l|l|l|l|l|l|l|l|l|l|l|l|l|}
\hline
Method & R@5 & R@10 & R@20 & R@50 & N@5 & N@10 & N@20 & N@50 & M@5 & M@10 & M@20 & M@50 \\
\hline
KGSF & 0.133 & 0.152 & 0.178 & 0.223 & 0.093 & 0.100 & 0.106 & 0.115 & 0.080 & 0.083 & 0.085 & 0.086 \\
MESE & 0.110 & 0.142 & 0.207 & 0.317 & 0.079 & 0.089 & 0.106 & 0.128 & 0.068 & 0.072 & 0.077 & 0.081 \\
UniCRS & 0.195 & 0.238 & 0.309 & \underline{0.406} & 0.147 & 0.160 & 0.178 & 0.197 & 0.131 & 0.136 & 0.141 & 0.144 \\
DCRS & 0.105 & 0.168 & 0.219 & 0.344 & 0.071 & 0.091 & 0.104 & 0.129 & 0.060 & 0.068 & 0.072 & 0.076 \\
STEP & 0.176 & 0.223 & 0.273 & 0.371 & 0.127 & 0.142 & 0.155 & 0.174 & 0.110 & 0.117 & 0.120 & 0.123 \\
MSCRS & \underline{0.207} & \underline{0.277} & \underline{0.332} & 0.387 & \underline{0.155} & \underline{0.178} & \underline{0.191} & \underline{0.202} & \underline{0.137} & \underline{0.147} & \underline{0.151} & \underline{0.153} \\
DACRS & $\textbf{0.230}^{*}$ & $\textbf{0.308}^{*}$ & $\textbf{0.348}^{*}$ & $\textbf{0.473}^{*}$ & $\textbf{0.160}^{*}$ & $\textbf{0.185}^{*}$ & \textbf{0.195} & $\textbf{0.220}^{*}$ & \textbf{0.139} & \textbf{0.151} & \textbf{0.155} & \textbf{0.158} \\
\hline
Improv. & \textcolor{ForestGreen}{11.11\%} & \textcolor{ForestGreen}{11.19\%} & \textcolor{ForestGreen}{4.82\%} & \textcolor{ForestGreen}{16.50\%} & \textcolor{ForestGreen}{3.23\%} & \textcolor{ForestGreen}{3.93\%} & \textcolor{ForestGreen}{2.09\%} & \textcolor{ForestGreen}{8.91\%} & \textcolor{ForestGreen}{1.46\%} & \textcolor{ForestGreen}{2.72\%} & \textcolor{ForestGreen}{2.65\%} & \textcolor{ForestGreen}{3.27\%} \\
\hline
\end{tabular}
\label{table:inspired rec}
\end{center}
\end{table*}

\subsection{Evaluation Metric}

To evaluate the recommendation performance of DACRS, we adopt Recall, NDCG (Normalized Discounted Cumulative Gain), and MRR (Mean Reciprocal Rank) as the evaluation metrics. Specifically, Recall@k measures how often the ground-truth item appears within the top-k recommended items.
\begin{equation}
    \text{Recall}@k = \frac{1}{N_t} \sum_{i=1}^{N_t} \mathbb{I}\left( y_i \in \hat{R}_i^{(k)} \right)
\end{equation}
where $N_t$ is the total number of test samples, $y_i$ is the ground-truth item for the $i$-th test sample, $\hat{R}_i^{(k)}$ is the top-k recommended item list for the $i$-th sample, and $\mathbb{I}(\cdot)$ is the indicator function, which returns 1 if the condition inside is true, and 0 otherwise.

We further evaluate recommendation ranking quality using NDCG and MRR. NDCG@k assigns higher scores to correctly recommended items that appear at higher ranks:
\begin{equation}
    \text{NDCG}@k = \frac{1}{N_t} \sum_{i=1}^{N_t} \frac{1}{\log_2(r_i + 1)}
\end{equation}
where $r_i$ denotes the rank position of the ground-truth item within the top-k recommendation list for the $i$-th sample. If the ground-truth item does not appear in the top-k list, the corresponding score is set to 0.

MRR@k evaluates the reciprocal rank of the first relevant item in the recommendation list:
\begin{equation}
    \text{MRR}@k = \frac{1}{N_t} \sum_{i=1}^{N_t} \frac{1}{r_i}
\end{equation}
where $r_i$ is the rank position of the ground-truth item for the $i$-th sample. Similarly, if the ground-truth item is not included in the top-k recommendations, the score is set to 0.

We do not evaluate response generation in this work, as our primary focus is on recommendation accuracy. Generating high-quality responses can be effectively handled by an LLM agent provided with the dialogue context and the recommended items \cite{huang2023recommender,feng2023large}.

\subsection{Baselines}
We compare DACRS with these SOTA CRS baselines:
\begin{itemize}
    \item \textbf{KGSF} \cite{zhou2020improving} builds a word KG and an entity KG and aligns the two latent embedding spaces through mutual information maximization,
    \item \textbf{MESE} \cite{yang2021improving} obtains item embeddings by encoding item textual metadata. It uses prompt learning for item retrieval.
    \item \textbf{UniCRS} \cite{wang2022towards} adopts prompt learning to generate embedding of a prompt token to retrieve items.
    \item \textbf{DCRS} \cite{dao2024broadening} retrieves similar dialogue contents as in-context demonstrations to enhance the recommendation.
    \item \textbf{STEP} \cite{yang2025step} uses curriculum learning to align the dialogue context with knowledge graph entities.
    \item \textbf{MSCRS} \cite{wei2025mscrs} incorporates multimodal information to construct modality-specific graphs and integrate them with prompt learning.
\end{itemize}

\subsection{Experimental Settings}
\label{appendix:setting}

For all baselines, we use their official implementations whenever available and evaluate them under the same experimental settings for fair comparison. We report the average results over five runs.

Following previous CRS works \cite{chen2019towards,zhou2020improving,wang2022towards}, we construct training samples by treating each utterance that contains at least one entity as a target instance. The previous dialogue context together with previously mentioned entities are used as model inputs, while the entities appearing in the current utterance are treated as ground-truth targets. We use the entity annotations provided by the datasets. During inference, we only evaluate system utterances that contain recommended items, following standard CRS evaluation protocols. Only item entities are considered as valid recommendation targets, and ranking is performed over all item entities in the dataset. For consistency, all compared methods are evaluated under the same protocol. We note that some previous works, such as DCRS, STEP, and MSCRS, additionally include user utterances during evaluation, which may lead to slight differences from their originally reported results.

For DACRS, we use Llama-3.2-3B-Instruct for dialogue rephrasing and summarization in the Dialogue Augmentation module. For dialogue encoding, we adopt GritLM-7B \cite{muennighoff2024generative} due to its strong retrieval-oriented representation capability. We use mean pooling over the final hidden states to obtain dialogue representations and keep the encoder frozen during training. Following prior CRS works \cite{chen2019towards,zhou2020improving,wang2022towards}, we employ a single-layer RGCN to encode the knowledge graph. The dialogue-guided attention network consists of one multi-head attention layer followed by a feed-forward layer.

During training, each dialogue is dynamically augmented online at every epoch. For both augmentation stages, one augmentation strategy (including No Augmentation) is randomly selected with equal probability.

We train DACRS using the AdamW optimizer with learning rate 0.001, weight decay 0.01, and batch size 128 for 50 epochs. All experiments are conducted on a single NVIDIA RTX A5000 GPU.

\subsection{RQ1: Recommendation Performance}
\label{sec:rq1}

\autoref{table:redial rec} and \autoref{table:inspired rec} report the recommendation performance on ReDial and Inspired, respectively. Overall, DACRS consistently achieves the best performance across almost all metrics on both datasets, demonstrating the effectiveness of the proposed framework in alleviating dialogue and item sparsity in conversational recommendation.

Specifically, on ReDial, DACRS improves Recall@10 from 0.230 to 0.255 and Recall@50 from 0.419 to 0.467 compared with the strongest baseline, corresponding to relative improvements of 10.87\% and 11.46\%, respectively. Similar improvements are consistently observed on ranking-sensitive metrics such as NDCG and MRR. For example, DACRS achieves an 8.91\% improvement on MRR@10. These results indicate that DACRS not only retrieves more relevant items, but also ranks target items at higher positions in the recommendation list.

On Inspired, DACRS achieves even larger gains. In particular, Recall@50 is improved by 16.50\% over the strongest baseline, while NDCG@50 is improved by 8.91\%. The improvements on Inspired are generally larger than those on ReDial, suggesting that the proposed framework is especially effective in relatively sparse scenarios with limited training data. This observation is consistent with our motivation that dialogue sparsity and item sparsity are key bottlenecks in conversational recommender systems.

We further observe that DACRS consistently improves not only Recall, but also NDCG and MRR across different cutoff values. Compared with Recall, NDCG and MRR place greater emphasis on the ranking positions of relevant items. The consistent gains on these metrics suggest that DACRS learns more discriminative user and item representations, enabling the model to rank user-preferred items more accurately rather than merely retrieving them within a large candidate set.

Among the baselines, KGSF achieves relatively weaker performance, which indicates the limitation of relying solely on conventional graph representation learning without leveraging the semantic understanding capability of LLMs. Meanwhile, recent LLM-enhanced CRSs such as UniCRS, STEP, and MSCRS achieve stronger performance by better utilizing dialogue semantics and knowledge graphs. However, these methods still suffer from dialogue variation and sparse entity supervision. In contrast, DACRS explicitly addresses both issues through dialogue augmentation and knowledge-guided entity modeling. The Dialogue Augmentation module improves robustness to diverse language expressions and reduces the influence of redundant textual patterns, while the Knowledge-Guided Entity Modeling module strengthens correlations among sparsely supervised entities through KG-based entity substitution and entity similarity regularization. Consequently, DACRS is able to learn more robust dialogue representations and more informative entity embeddings, leading to consistent improvements across different datasets and evaluation metrics.

\begin{figure}[t]
\centering
    \includegraphics[width=0.48\textwidth]{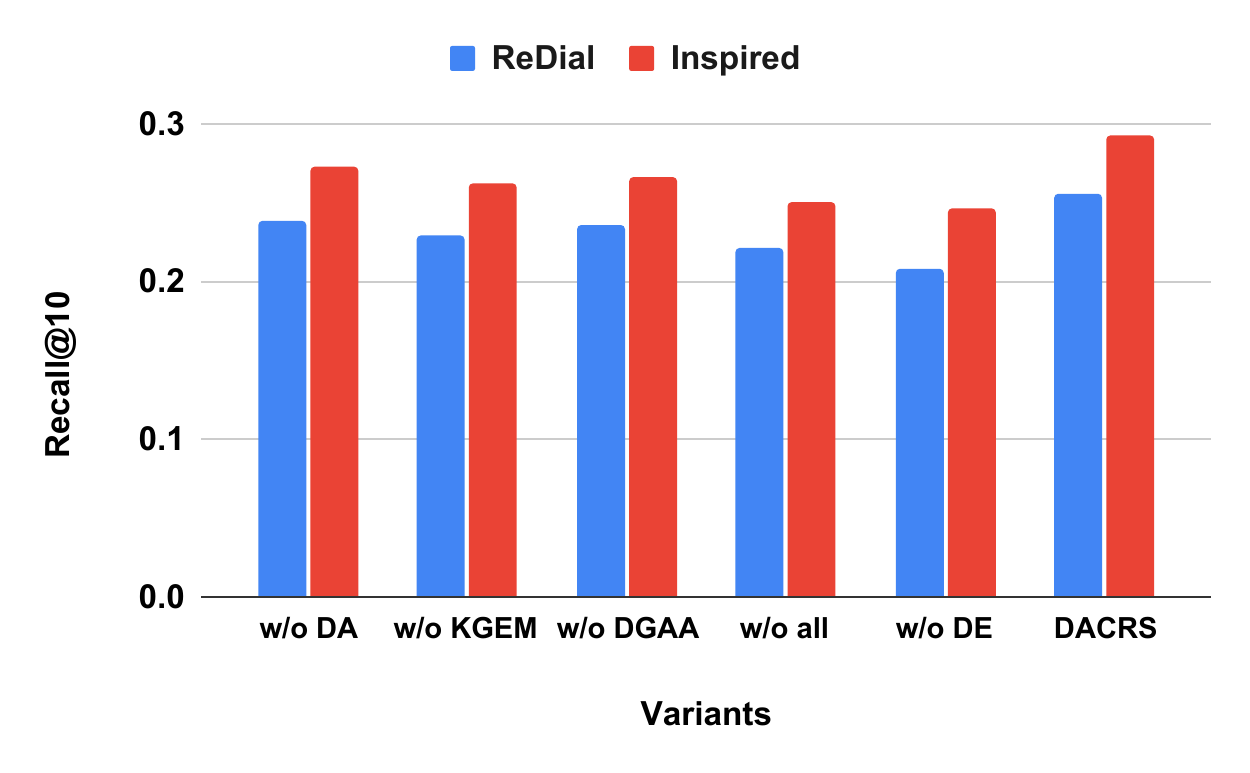}
\caption{Ablation study of DACRS. Each component contributes positively to the overall performance. Eliminating dialogue context results in the most severe degradation, underscoring its necessity for preference modeling, while removing KGEM leads to the largest drop among dialogue-aware variants, highlighting the importance of knowledge-guided entity modeling for capturing item correlations.}
\label{fig:ablation}
\end{figure}

\begin{figure*}[t]
\centering
    \subfloat[Loss weight $\alpha$.]{\includegraphics[width=0.33\linewidth]{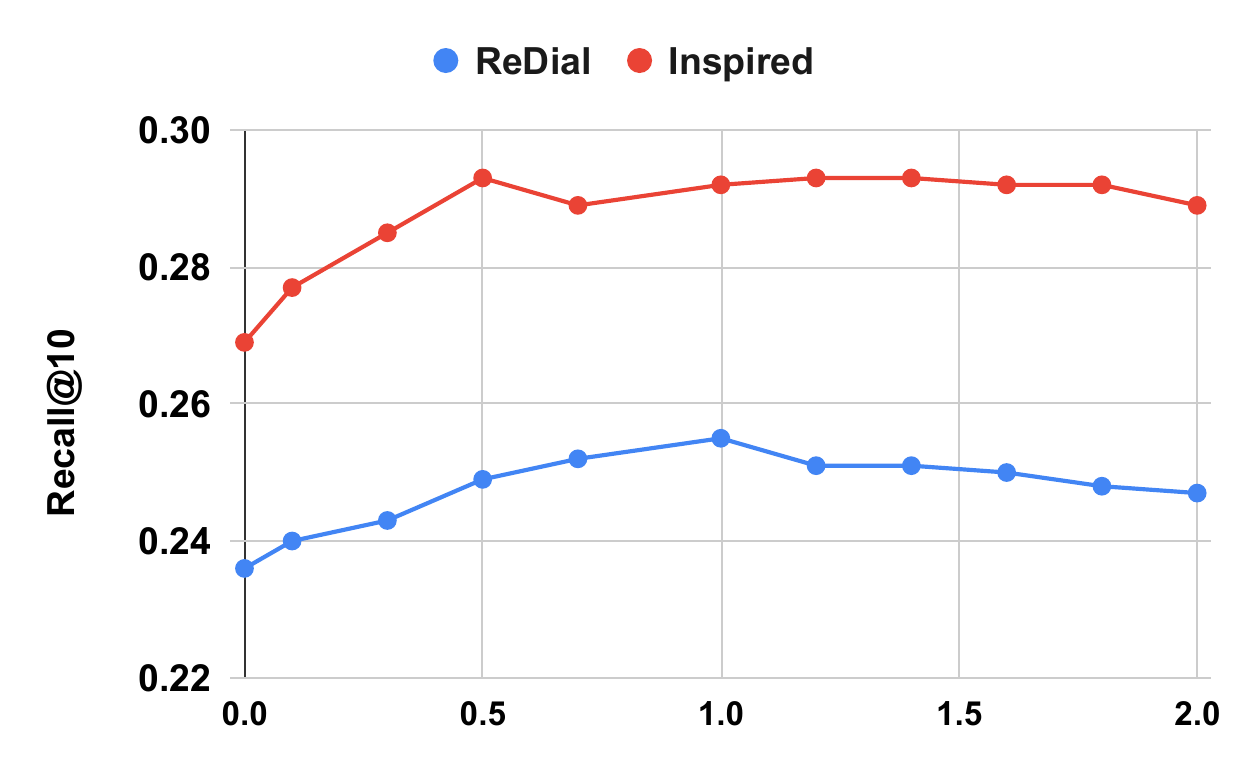}\label{fig:param}}
	\subfloat[Entity substitution rate.]{\includegraphics[width=0.33\linewidth]{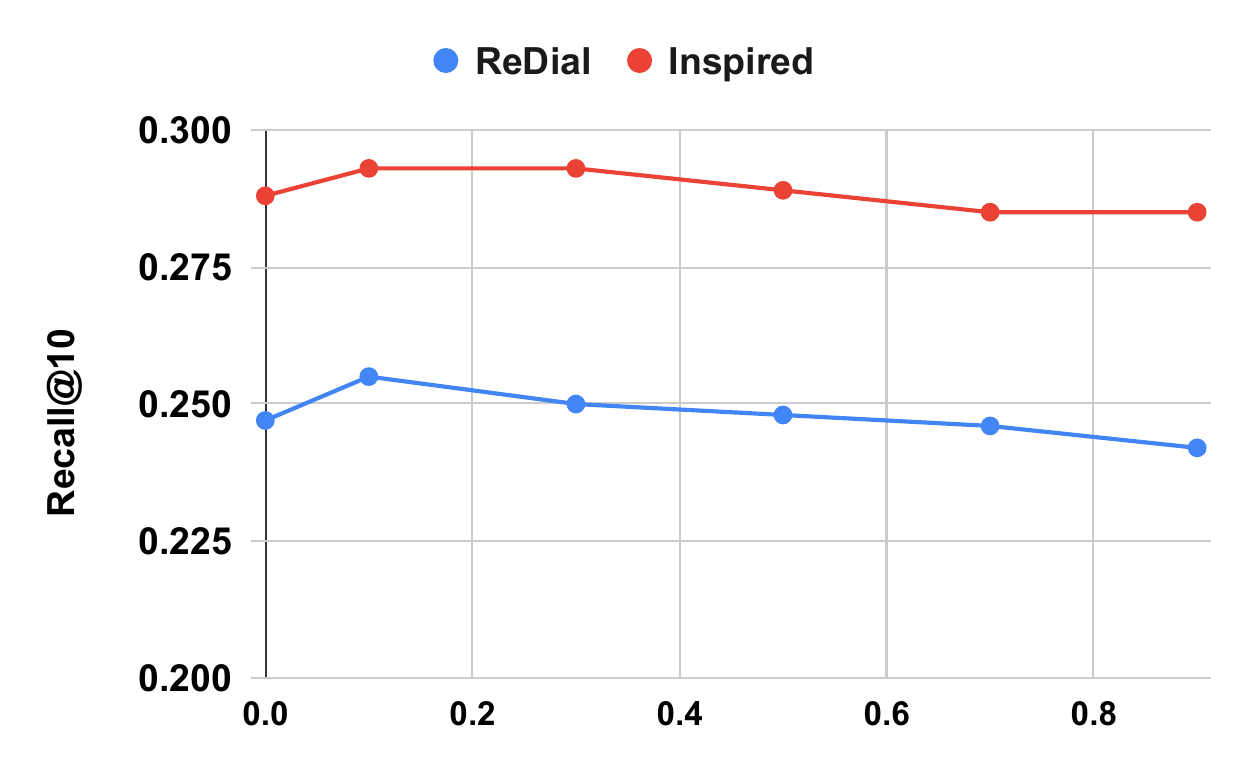}\label{fig:sub}}
    \subfloat[Dialogue augmentation rate.]{\includegraphics[width=0.33\linewidth]{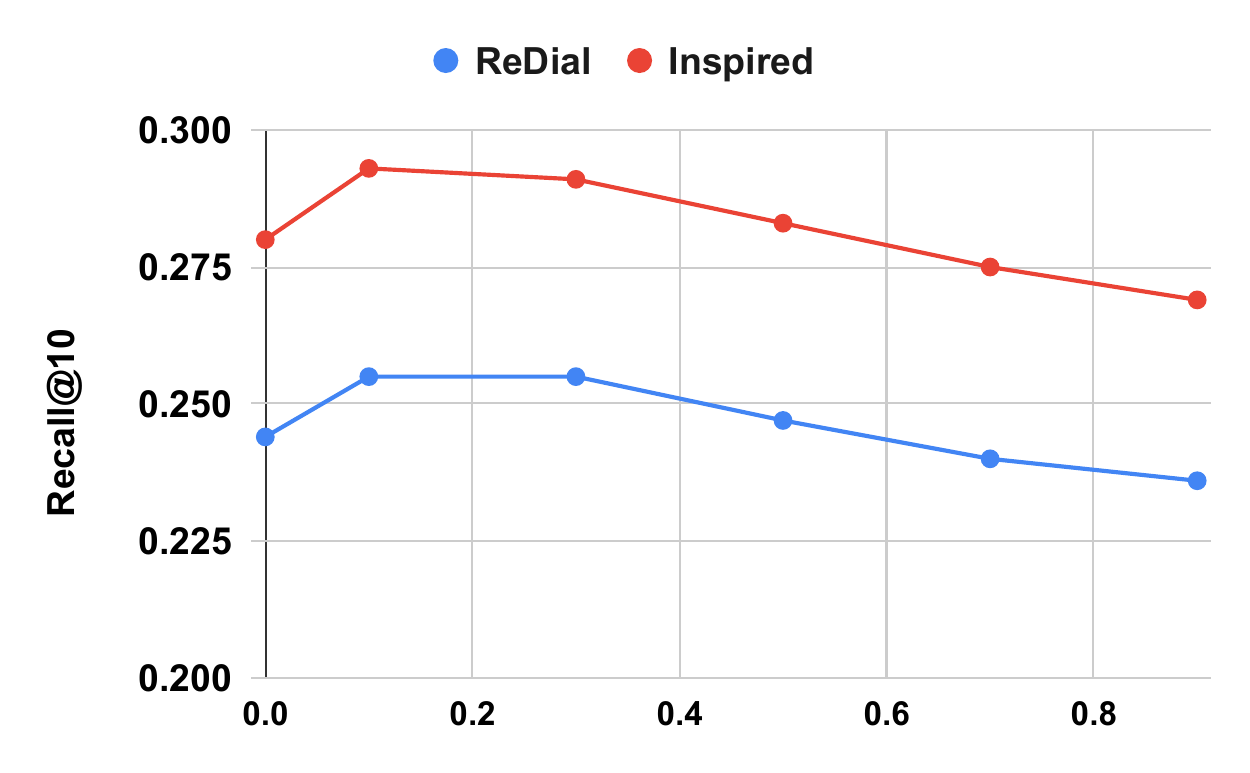}\label{fig:aug}}
\caption{Hyperparameter study of DACRS.
The figure illustrates the sensitivity of DACRS to three key hyperparameters: (a) the loss weight $\alpha$, which shows stable performance within a reasonable range; (b) the entity substitution rate, where moderate substitution is beneficial while excessive substitution introduces noise; and (c) the dialogue augmentation rate, where mild augmentation improves generalization but overly strong augmentation causes information loss.}
\label{fig:param study}
\end{figure*}

\subsection{RQ2: Ablation Study}
\label{sec:rq2}
We test how each module contributes to our method in \autoref{fig:ablation}. We test these variants: 
\begin{itemize}
    \item w/o DA: remove the Dialogue Augmentation module and use the vanilla dialogue to get dialogue embeddings.
    \item w/o KGEM: remove the Knowledge-Guided Entity Modeling module such that the entity embeddings are directly acquired from the RGCN without any augmentations.
    \item w/o DGAA: replace the Dialogue-Guided Attention Aggregation module with a dot product between the dialogue embedding and the candidate item embeddings. 
    \item w/o all: remove all above three components.
    \item w/o DE: remove DA and DGAA. Recommendation is made by the dot product between the average embedding of the mentioned entities and the item embeddings.
\end{itemize}
From \autoref{fig:ablation}, we have several important observations.

First, all ablated variants consistently underperform DACRS on both datasets, demonstrating that each proposed component contributes positively to the final recommendation performance. This validates the overall design of DACRS, where dialogue augmentation, knowledge-guided entity modeling, and dialogue-entity matching work collaboratively to alleviate different forms of sparsity in CRS.

Second, removing the dialogue-related components leads to the most severe performance degradation. In particular, w/o DE completely eliminates dialogue semantic modeling and only relies on mentioned entity embeddings for recommendation, resulting in the worst performance among all variants. This observation highlights that dialogue context is the primary source for modeling user preference in CRS, since conversational semantics often contain explicit fine-grained preference signals that cannot be captured solely by mentioned entities.

Third, among the variants that still preserve dialogue information, w/o KGEM exhibits the largest performance drop. This result demonstrates the importance of addressing entity sparsity and weak entity correlations in conversational recommendation. Without knowledge-guided entity modeling, many entities receive limited supervision and fail to learn informative representations, especially under long-tail and sparse KG settings. In contrast, KG-based entity substitution and the entity similarity constraint enable DACRS to propagate informative signals across connected entities, resulting in more robust entity representations.

Finally, w/o DGAA also leads to noticeable degradation, indicating that directly matching dialogue embeddings with item embeddings is insufficient for accurate recommendation. The dialogue-guided attention aggregation mechanism effectively fuses explicit semantic preferences from dialogue texts with implicit collaborative signals from mentioned entities, producing more informative user representations for recommendation.

\subsection{RQ3: Hyperparameter Study}
\label{appendix:hyperparameter}
\subsubsection{Loss Weight $\alpha$.}

As shown in Figure \ref{fig:param}, the performance first improves as $\alpha$ increases from 0 to 0.5, indicating that the entity similarity constraint effectively enhances representation learning by encouraging structurally related entities to share similar embeddings. This regularization alleviates the sparsity issue for rarely supervised entities and improves the generalization ability of the model.

However, when $\alpha$ becomes excessively large, the performance slightly decreases. This suggests that overemphasizing structural similarity may weaken the model's ability to preserve discriminative signals among different entities, leading to oversmoothed entity representations. Therefore, a moderate constraint strength achieves the best balance between structural regularization and recommendation discrimination.

\subsubsection{Entity Substitution Rate.}
 
Figure \ref{fig:sub} shows that a moderate substitution rate achieves the best performance, while overly aggressive substitution leads to performance degradation. This phenomenon reflects a trade-off between entity diversity and semantic consistency.

A moderate amount of KG-based substitution exposes the model to a broader range of semantically related entities, which improves the generalization ability of entity representations under sparse supervision. However, excessive substitution may distort the original user preference semantics, since neighboring entities in the KG are not always perfectly aligned with the user's actual interests. Consequently, overly large substitution rates introduce semantic drift and noise into preference modeling.

\subsubsection{Dialogue Augmentation Rate.}

As shown in Figure \ref{fig:aug}, moderate dialogue augmentation consistently improves recommendation performance, demonstrating that augmentation helps the model generalize to diverse linguistic expressions and reduces sensitivity to redundant textual patterns.

However, when the augmentation rate becomes excessively large, the performance begins to decline. Heavy augmentation may substantially alter the original dialogue semantics or remove critical preference signals, resulting in a mismatch between augmented training data and real dialogue distributions during inference. Therefore, mild augmentation is more effective for improving robustness while preserving semantic fidelity.

\begin{figure}[t]
\centering
    \subfloat[w/o KGEM]{\includegraphics[width=0.5\linewidth]{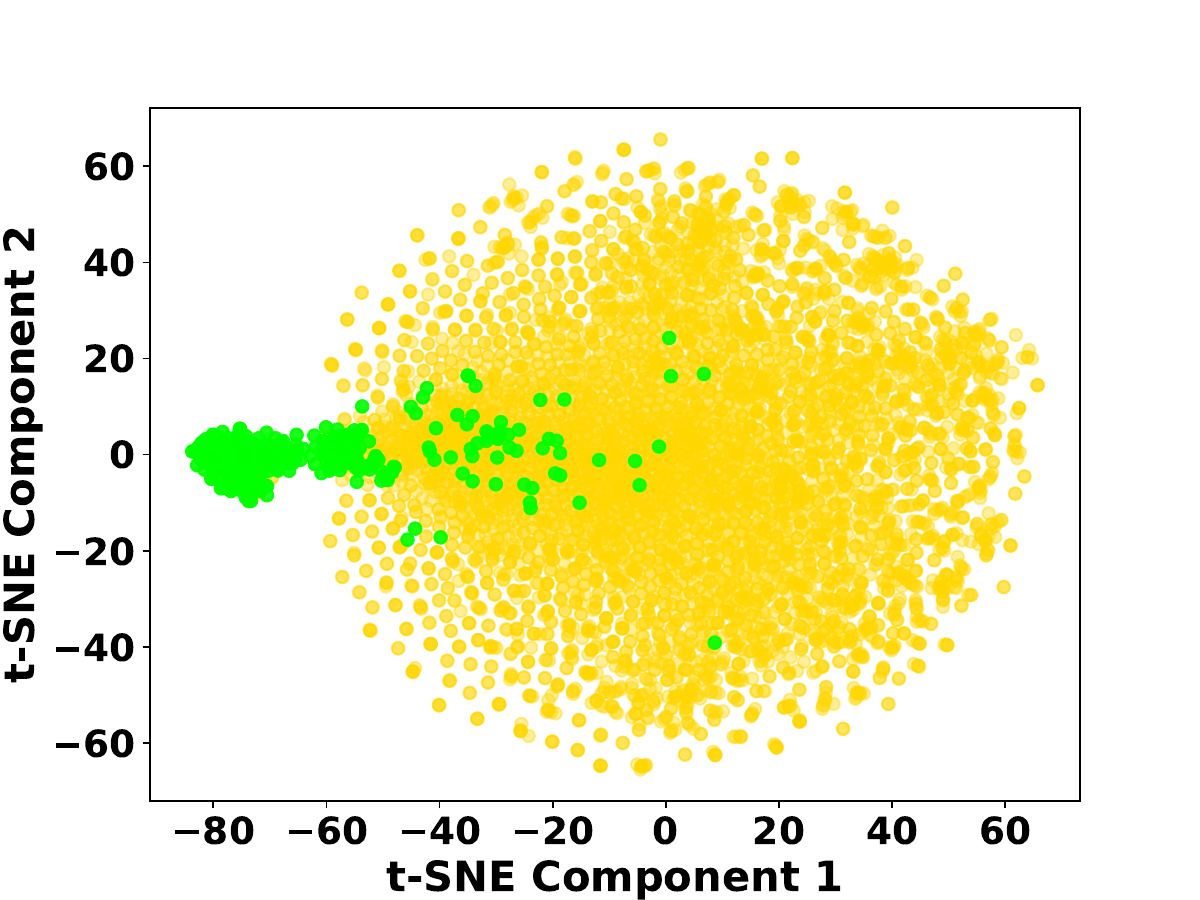}\label{fig:wokgem}}
	\subfloat[with KGEM]{\includegraphics[width=0.5\linewidth]{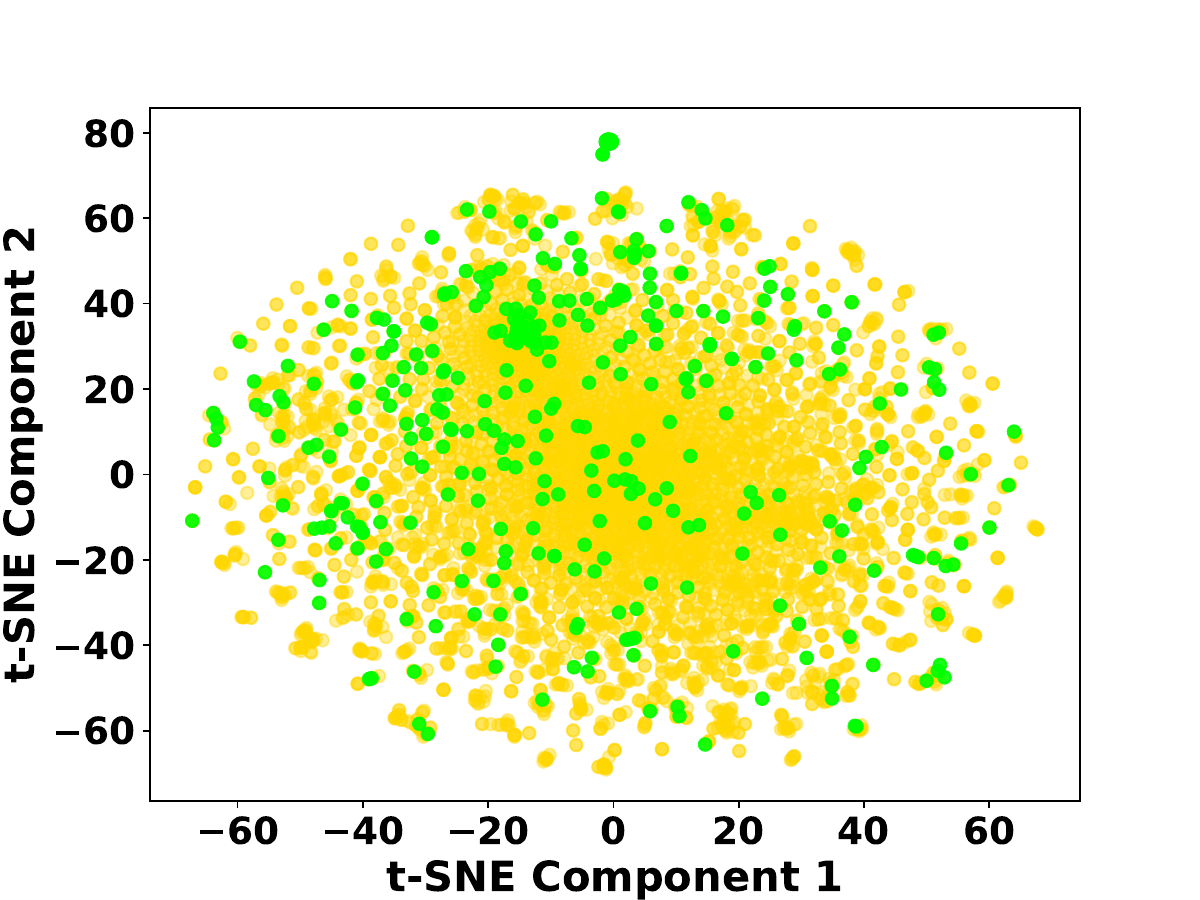}\label{fig:wkgem}}

\caption{Visualization of ReDial entity embeddings with and without Knowledge-Guided Entity Modeling. Green nodes represent test-only entities.}
\label{fig:visualization}
\end{figure}

\subsection{RQ4: Visualization of Entity Embeddings}
\label{sec:visualize}

To better understand how the proposed Knowledge-Guided Entity Modeling (KGEM) module influences entity representation learning, we visualize the entity embeddings learned with and without KGEM using t-SNE, as shown in \autoref{fig:visualization}. In the visualization, entities that appear only in the test set but not in the training set are highlighted in green, while all remaining entities are shown in yellow.

The left plot presents the embedding space learned by the baseline model without KGEM. A clear pattern is that the test-only entities deviate from the main embedding distribution and form a relatively isolated cluster. This indicates that, without KGEM, entities that receive little or no direct supervision during training are not well aligned with the majority of learned entity representations. Since these entities rarely participate in user-item interactions or effective knowledge-guided updates, their embeddings tend to remain under-trained and poorly integrated into the overall representation space. Such separation makes it difficult for the model to compare these entities with well-trained entities in a meaningful way, which may hurt recommendation performance especially for sparse or unseen entities.

In contrast, the right plot shows the embedding distribution when KGEM is incorporated. The previously isolated cluster largely disappears, and the test-only entities become mixed with the yellow entities in the main distribution. This suggests that KGEM can propagate useful information through knowledge-guided relations, allowing entities with limited direct supervision to receive indirect training signals from related entities. As a result, these sparsely supervised entities are better integrated into the learned embedding space instead of being pushed into a separate region.

From a recommendation perspective, this improved integration is important because the model can represent sparse or test-only entities in a space that is more consistent with the rest of the entity embeddings. This makes it easier to distinguish relevant entities from irrelevant ones and provides a plausible explanation for the consistent improvements in Recall, NDCG, and MRR observed in the main experiments.

Overall, the visualization provides qualitative evidence that KGEM alleviates the entity sparsity issue by improving the representation learning of entities that would otherwise remain under-trained or poorly aligned with the main entity embedding space.